\documentclass[12pt]{iopart}
\usepackage{cite}

\usepackage{graphicx}
\usepackage{subfigure}
\usepackage{array}
\usepackage[colorlinks,
    linkcolor=blue,
    anchorcolor=blue,
    citecolor=blue
]{hyperref}
\usepackage{soul}
\soulregister\cite7
\soulregister\citep7
\soulregister\citet7
\soulregister\ref7
\soulregister\pageref7
\begin{document}

\title{Characterization of Pedestal Burst Instabilities during I-mode to H-mode Transition in the EAST Tokamak}

\author{X.M. Zhong$^{1}$, X.L. Zou$^{2}$, A.D. Liu$^{*1}$, Y.T. Song$^{*3}$, G. Zhuang$^{1}$, E.Z. Li$^{3}$, B. Zhang$^{3}$, J. Zhang$^{1}$, C. Zhou$^{1}$, X. Feng$^{4}$, Y.M. Duan$^{3}$, R. Ding$^{3}$, H.Q. Liu$^{3}$, B. Lv$^{3}$, L. Wang$^{3}$, L.Q. Xu$^{3}$, L. Zhang$^{3}$, Hailin Zhao $^{3}$, Qing Zang $^{3}$, Tao Zhang $^{3}$, B.J. Ding$^{3}$, M.H. Li$^{3}$, C.M. Qin$^{3}$, X.J. Wang$^{3}$, X.J. Zhang$^{3}$ and EAST Team$^{3}$}
\address{$^{1}$School of Nuclear Science and Technology, University of Science and Technology of China,
Anhui Hefei 230026, China
\\
$^2$ CEA, IRFM, F-13108 St Paul Les Durance, France
\\
$^{3}$Institute of Plasma Physics, Chinese Academy of Sciences, Anhui Hefei 230021, China
\\
$^{4}$Shenzhen University, Guangdong Shenzhen 518061, China}
\ead{lad@ustc.edu.cn, songyt@ipp.ac.cn}
\vspace{10pt}
\begin{indented}
\item[]\today
\end{indented}

\begin{abstract}
Quasi-periodic Pedestal Burst Instabilities (PBIs), featuring alternative turbulence suppression and bursts, have been clearly identified by various edge diagnostics during I-mode to H-mode transition in the EAST Tokamak.
The radial distribution of the phase perturbation caused by PBI shows that PBI is localized in the pedestal.
Prior to each PBI, a significant increase of density gradient close to the pedestal top can be clearly distinguished, then the turbulence burst is generated, accompanied by the relaxation of the density profile, and then induces an outward particle flux.
The relative density perturbation caused by PBIs is about $6- 8\%$. 
Statistic analyses show that the pedestal normalized density gradient $R/L_n$ triggering the first PBI has a threshold value, mostly in the range of $22 - 24$, suggesting that a PBI triggering instability could be driven by the density gradient.
And $R/L_n$ triggering the last PBI is about $30 - 40$ and seems to increase with the loss power and the chord-averaged density. 
In addition, the frequency of PBI is likely to be inversely proportional to the chord-averaged density and the loss power.
These results suggest that PBIs and the density gradient prompt increase prior to PBIs can be considered as the precursor for controlling I-H transition.
\\
\item Key words: I-mode, I-H transition, pedestal burst instabilities, density gradient
\end{abstract}

\section{Introduction}

High confinement mode (H-mode) features edge transport barrier (ETB) with steep density and temperature gradients in the pedestal region due to the turbulence suppression \cite{wagner1982,wagner1984}.
And H-mode has been considered as the baseline operation scenario for the International Thermonuclear Experimental Reactor (ITER) \cite{rebut1995}. 
However, repetitive magnetohydrodynamic (MHD) instabilities, referred to as edge-localized modes (ELMs), which are due to the relaxation of edge pressure and current, produce high transient heat loads on the plasma facing components (PFCs) \cite{zohm1996}. 
For the Type-I ELMs, the peak heat load may exceed the PFCs tolerable limit, which is unacceptable \cite{leonard2014edge}. 
Thus, the ELMs mitigation and suppression is one of the main research topics in magnetic confinement fusion.

Now, the methods of ELMs mitigation and suppression can be roughly divided into two types, namely, external actuators and scenario based  method by searching small/no ELMs regimes. 
External actuators include resonant magnetic perturbation (RMP) \cite{evans2006edge}, pellet injection (PI) \cite{milora1995pellet}, supersonic molecular beam injection (SMBI) \cite{xiao2012elm,zheng2013comparison}, LHCD \cite{xiao2017effect}, impurity \cite{zhang2018control} and so on.
Another way is to search small/no ELMs regimes, such as  QH-mode \cite{burrell2001quiescent}, EDA H-mode \cite{greenwald1999characterization}, HRS H-mode \cite{kamiya2003observation}, Grassy-ELM \cite{kamada2000disappearance}, Type II-ELM \cite{stober2001type}, Type V-ELM \cite{maingi2005observation}, and I-mode \cite{walk2014edge,Happel2016PPCF}.
The existence of particle transport barrier in H-mode plasma is not favorable for avoiding impurity accumulation in the plasma core and helium ash removal.
I-mode, featuring high energy confinement comparable to H-mode and moderate particle confinement comparable to L-mode \cite{whyte2010mode}, can be a potential candidate for future fusion devices due to the lack of ELMs.
As a consequent, I-mode is also losing the benign effect impurity flushing of the ELMs.
In addition, it should be mentioned that, the power at the transition from L-mode to I-mode in the unfavorable configuration $P_{L-I}$ is clearly higher than the usual L-H threshold power in the favorable configuration $P_{L-H}$ in high density plasmas \cite{ryter2016mode}.
Similar results can also be found in EAST I-mode discharges \cite{liu2020power}, suggesting that currently I-mode could be difficult to obtain in ITER and decreasing the power threshold of L-I transition is the key issue for the future I-mode research.

I-mode was reported firstly in Alcator C-Mod  (C-Mod)\cite{greenwald1997h} and ASDEX Upgrade (AUG) \cite{Ryter_1998}.
Generally, I-mode is usually obtained in the unfavorable configuration, i.e. the  $B$$\times$$\nabla$$B$  ion drift pointing away from the active X-point.
In C-Mod and AUG, I-mode is always accompanied by the weakly coherent mode (WCM) and the geodesic-acoustic mode (GAM) \cite{whyte2010mode,manz2015geodesic}.
The WCM, which is localized at the pedestal and is considered as a signature, is believed to be responsible for no particle transport barrier in I-mode \cite{Manz2020NF,zxliu2016pop}.
In C-Mod, it has been found that the access of I-mode in unfavorable configuration is easier than that in favorable configuration.
This geometric asymmetry is explained by the nonlinear turbulence interactions with/without GAM \cite{Cziegler2017PRL}.  

Recently, the stationary I-mode regime has been identified in EAST \cite{Liu_2020, feng2019mode}. 
Similar with other devices, I-mode is always accompanied by the WCM with the frequency range of $40 - 150 kHz$ in the pedestal. 
In addition, a low-frequency coherent mode of $6 - 12 kHz$, which is identified as a radially localized edge temperature ring oscillation (ETRO) \cite{Liu_2020}, is always concomitant. 
Furthermore, it has been found that the ETRO is probably caused by the alternating transitions between two kinds of turbulence in ion and electron diamagnetic drift direction.
It should be noticed that, GAM is often absent where the WCM is most significant in EAST \cite{feng2019mode}. 

However, it has been observed that the stationary I-mode can transit into the H-mode spontaneously or with the increase of auxiliary heating power. 
To maintain the stationary I-mode, identifying the precursor and actuator for I-H transition is a critical issue in the magnetic confinement fusion research.
Here, instabilities with periodic burst have been observed during the I-H transition in EAST. 
These burst instabilities are localized in the pedestal.
Therefore, we refer to them as pedestal burst instabilities (PBIs) in the following. 
PBIs can be observed by various diagnostics, such as electron cyclotron emission (ECE), Doppler Reflectometer (DR), Bolometer, Soft X-Ray (SXR), Mirnov probes and Divertor Langmuir probes (Div-LPs).
Similar events were reported in C-Mod \cite{walk2014edge, Silvagni2021NF}, DIII-D \cite{hubbard2016multi} and AUG \cite{Silvagni_2020}.
These events are called Pedestal Relaxation Events (PREs) in both AUG and C-Mod \cite{Silvagni2021NF}.
However, the characteristics of these quasi-periodic instabilities are still not clear.
In this work, the characterization of PBIs will be discussed in details.

The paper is organized as follows. 
Section 2 presents experimental condition and diagnostics.
In Section 3, the I-H transition experimental results are reported. 
And the identification, characteristics and statistic results of PBIs are shown in Section 4. 
Finally, Section 5 is the conclusion.

\section{Experimental conditions and diagnostics}
I-H transition experiments have been performed in EAST, with the plasma major radius $R = 1.9 m$ and the plasma minor radius $a = 0.45 m$ \cite{wu2007overview,wan2017overview,li2013long}. 
The auxiliary heating in EAST include lower hybrid current drive (LHCD), electron cyclotron resonance heating (ECRH), ion cyclotron resonance heating (ICRH) and neutral beam injection (NBI).

The turbulence rotation and intensity are measured by the 8-channel DR \cite{Zhou2013RSI}.
The measured turbulence perpendicular wavenumber $k_\perp$ at the cutoff layer is $4 - 6 cm^{-1}$. 
And the Doppler shift $f_d$  of the turbulence can be expressed as $f_d$ = $(k_\perp$ $\ast$ $\upsilon$$_\perp) /(2\pi)$ \cite{zou1999eps}, among which $\upsilon$$_\perp$ is the turbulence perpendicular rotation velocity relative to the direction of the magnetic field. 

The DR phase derivative perturbation can be written as \cite{hillesheim2010rsi}: 

$$\frac{d\widetilde{\phi }}{dt}= k_\perp (  \widetilde{V}_{E\times B} + \widetilde{V}_{phase}  )  +  \frac{d\widetilde{\phi }_0}{dt} $$
where the first term on the right hand corresponds to the backscattering and the second term on the right hand corresponds to the phase modulation due to cut-off oscillation \cite{fanack1996ppcf}, where $\widetilde{V}_{E\times B} $ represents fluctuation of $E$$\times$$B$ velocity, $\widetilde{V}_{phase}$ is turbulence phase velocity fluctuation, and ${d\widetilde{\phi }_0}$ describes density fluctuation at the cut-off layer.
In I-mode, the WCM is measured with the phase modulation \cite{Zhong2016ppcf}, while the ETRO is measured by the turbulence phase velocity term $V_{phase}$ \cite{Liu_2020}, and the GAM, which is a high frequency zonal flow, can be observed via $E$$\times$$B$ velocity term in the $d\widetilde{\phi } / dt$ spectra \cite{feng2019mode,Liu_2020}.

The density is measured by 3-channel hydrogen cyanide (HCN) laser interferometer \cite{SHEN20132830}, 11-channel polarimeter-interferometer (POINT) \cite{Liu2014RSI} and multi-channel density profile reflectometer \cite{Xiang2018RSI}. Moreover, the electron temperature $T_e$ profile is measured by 32-channel ECE \cite{LIU201872} and Thomson scattering (TS) system \cite{qing2010development}. In addition, the plasma radiation is measured by 64-chords fast bolometer \cite{duan2012resistive} and SXR \cite{chen20162}. And the magnetic fluctuation is measured by Mirnov coils. The divertor particle flux is measured by Div-LPs \cite{Xu2016RSI}.

\section{I-H transition in EAST}

\begin{figure}[htbp]
\centering
\includegraphics[width=7in]{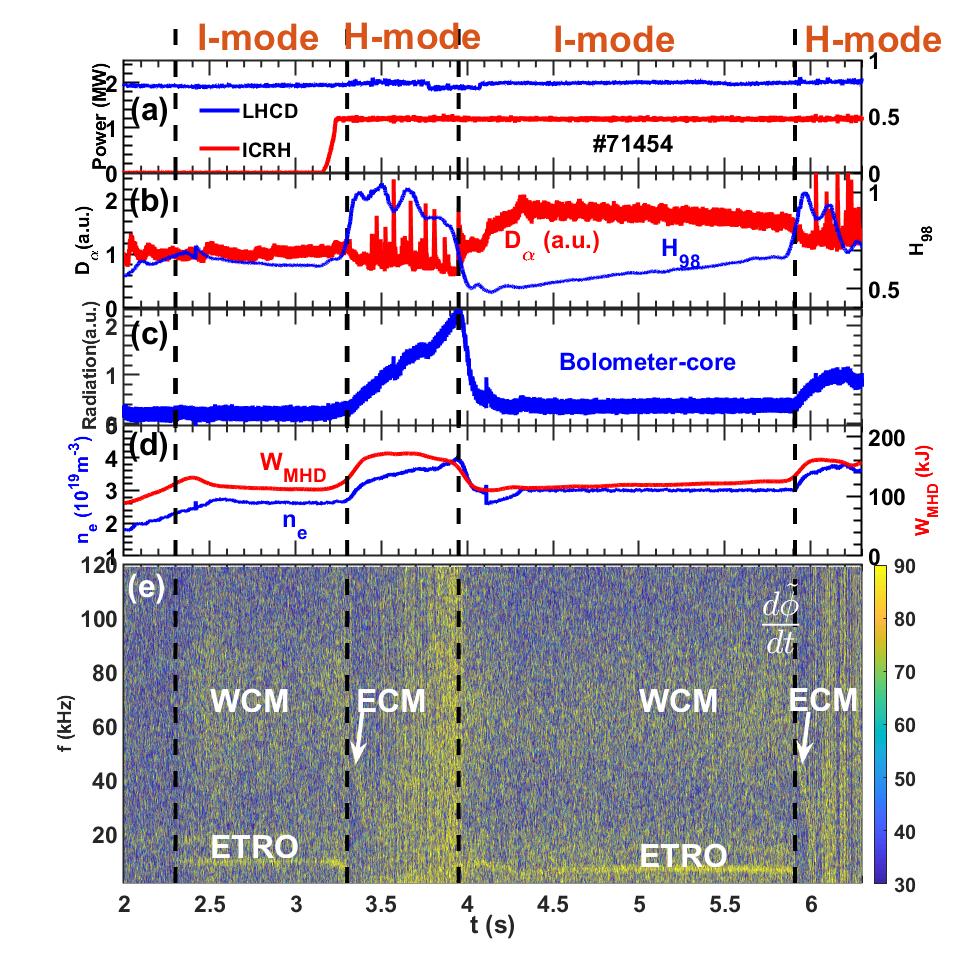}
\caption{I-H transition in the EAST typical discharge. (a) LHCD (blue line) and ICRH (red line) power. (b) $D_\alpha$ signal (red line) and {$H_{98}$ factor signal (blue line)}. (c) Bolometer signal in the core region. (d) Chord-averaged density (blue line) and plasma stored energy $W_{MHD}$ (red line). (e) Time-frequency spectra of the DR phase derivative signal}
\label{IH71454}
\end{figure}
Fig. \ref{IH71454} displays a typical I-H transition in EAST, with the plasma current  $I_p = 600 kA$  and the toroidal magnetic field $B_t = 2.4 T$.
The plasma is maintained under $2.0 MW$ LHCD, and  addition power of  $1.2 MW$ ICRH is injected  at $t = 3.15 s$  (Fig. \ref{IH71454}a).
The L-I transition at $t = 2.35 s$ can be clearly identified by the appearances of the WCM and the ETRO in the  $d\widetilde{\phi } / dt$ spectra, as well as the increase of the $H_{98}$ factor, while the edge $D_\alpha$ signal remains nearly unchanged.
ICRH triggers the H-mode at $t = 3.3 s$, which can be identified by the sudden decrease of the $D_\alpha$ signal and the following bursts caused by ELMs, as well as the distinct increase of the chord-averaged density, the plasma stored energy and the $H_{98}$ factor.
Meanwhile, the edge coherent mode (ECM) with frequency of $30 - 50 kHz$ appears in the $d\widetilde{\phi } / dt$ spectra, which is a common mode appearing at the H-mode pedestal region in EAST \cite{JI_2021}.
The H-mode is maintained till $t = 3.95 s$ and then transits back to I-mode due to the core impurity accumulation represented by the core radiation power, which can be seen from Fig. \ref{IH71454}c.
Finally another I-H transition occurs at $t = 5.91 s$.
And this may be due to an increase in the pedestal normalized density gradient, which will be discussed in details  later.

\begin{figure}[htbp]
\centering
%\subfigure{
%\label{Fig. sub.1}
%\includegraphics[width=3.58in]{ne75357.jpg}
%}
%\subfigure{
%\label{Fig. sub.2}
%\includegraphics[width=3.45in]{te75357.jpg}
%}
%\subfigure{
%\label{Fig. sub.3}
%\includegraphics[width=4in]{dphidt75357.jpg}
%}
\centering
\includegraphics[width=4in]{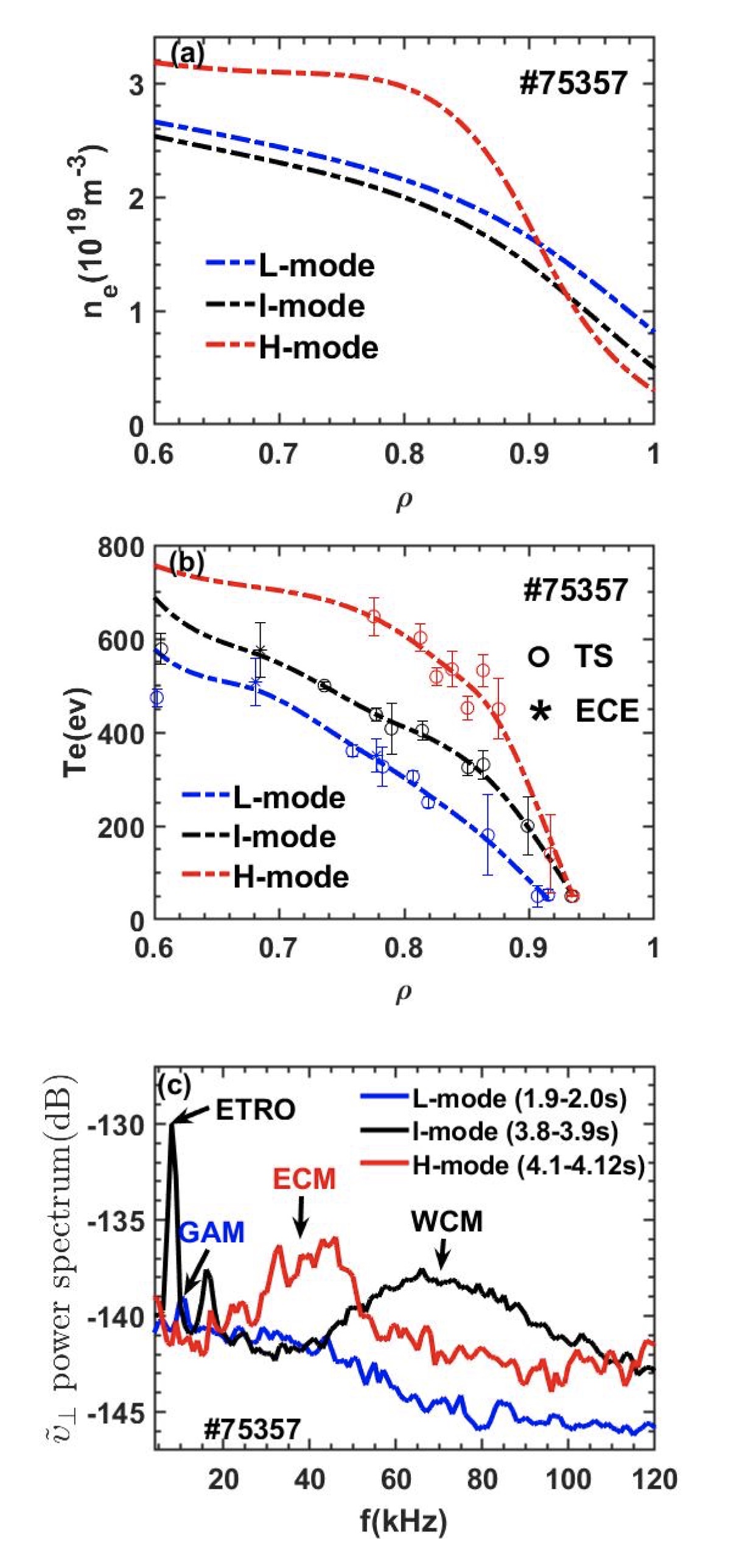}
\caption{(a) Density profile during L, I, and H-mode. (b) Electron temperature profile during L, I, and H-mode. (c) $d\widetilde{\phi } / dt$ power spectrum of DR during L, I, and H-mode.}
\label{LIH75357}
\end{figure}
The profiles of electron density and electron temperature during L, I, and H-mode are shown in Fig. \ref{LIH75357}a and Fig. \ref{LIH75357}b respectively.
The density profile in I-mode is similar with that in L-mode and is much lower than that in H-mode (Fig. \ref{LIH75357}a).
Fig. \ref{LIH75357}b displays that I-mode has a significant temperature edge pedestal.
The comparison of $d\widetilde{\phi } / dt$ power spectrum at the plasma edge region is shown in Fig. \ref{LIH75357}c.
The $13 kHz$ GAM can be clearly observed in L-mode, and the WCM with frequency of $40 - 100 kHz$ and the ETRO with frequency of 10 kHz can be found in I-mode.
{As shown in Fig. \ref{IH71454} and Fig. \ref{PBI71454} later, the frequency of the ECM decreases strongly after the onset due to the plasma rotation\cite{JI_2021}.
It can found that the WCM spectrum is much larger than that of the ECM, as shown in Fig. \ref{LIH75357}c.}

\section{Pedestal burst instabilities during I-H transition}
\subsection{Identification of pedestal burst instabilities}
\begin{figure}[htbp]
\centering
\includegraphics[width=6.8in]{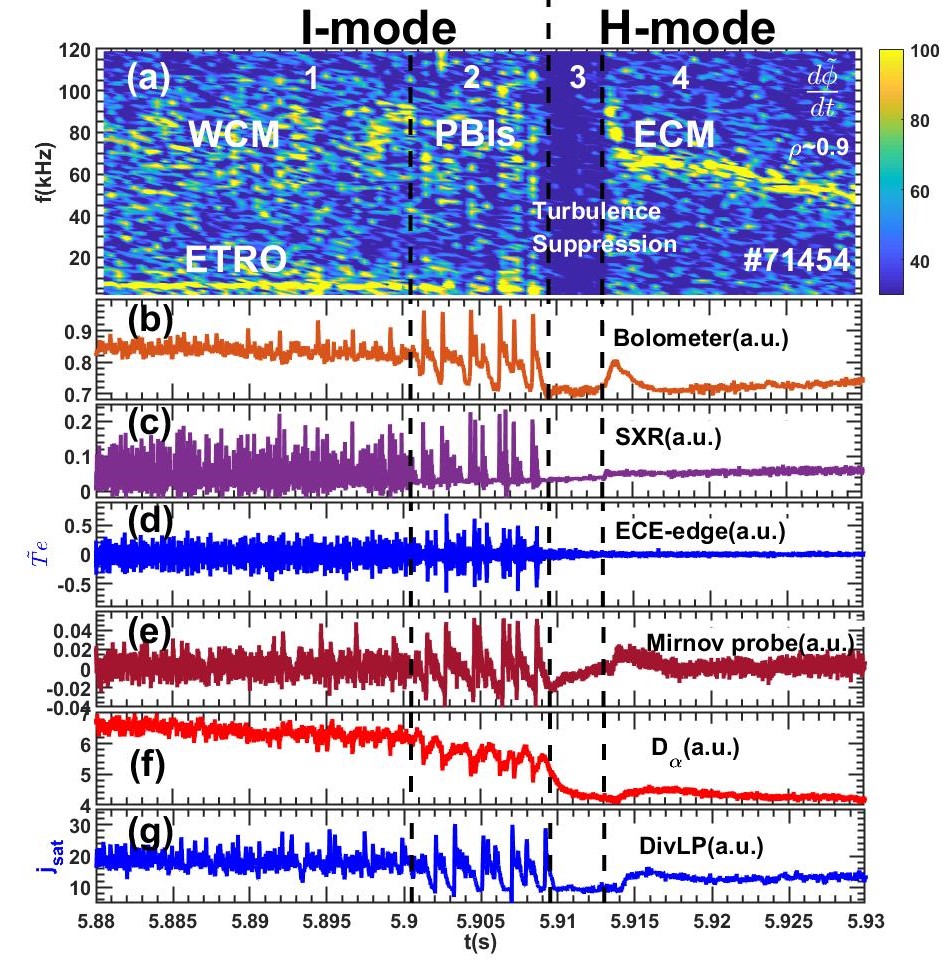}
\caption{Identification of PBIs by various diagnostics. (a) Frequency spectrogram of $d\widetilde{\phi } / dt$  at $\rho  = 0.9$. (b) Bolometer signal. (c) Soft X-ray signal. (d) Edge $T_e$ perturbation. (e) Mirnov probe signal. (f) $D_\alpha$ signal. (g) Div-LP signal.}
\label{PBI71454}
\end{figure}
Fig. \ref{PBI71454}a to Fig. \ref{PBI71454}g are the time-frequency spectrogram of $d\widetilde{\phi } / dt$ , the bolometer signal, the SXR signal, the temperature perturbation signal, the Mirnov probe signal, the $D_\alpha$ signal, and the Div-LPs signal, respectively.
It can be found that a quasi-periodic instability emerges during the I-H transition, and such instability can be observed by most of the edge diagnostics.
Considering that it acts like period bursts and is excited close to the temperature pedestal region in I-mode as shown later, these instabilities are referred as Pedestal Burst Instabilities (PBIs) in the following.

Furthermore, based on features of the signal evolution, the whole I-H transition process can be divided into four phases.
Interval 1 corresponds to the stationary I-mode, with the WCM and the ETRO. 
Interval 2 is the PBI-phase, during which the bursts can be identified by various diagnostics. 
In interval 3, H-mode appears, accompanying with a short interval, significant reductions of radiation and particle flux in the pedestal region, as well as the strong reduction of the turbulence intensity.
Therefore, this interval is called as the turbulence suppression phase.
Interval 4 is the H-mode phase with ECM. 
The first ELM appears at about $t = 6.02 s$, which is not shown here and can be seen from Fig. \ref{IH71454}b.

\subsection{Radial localization of PBI}

\begin{figure}[htbp]
\centering
\includegraphics[width=7in]{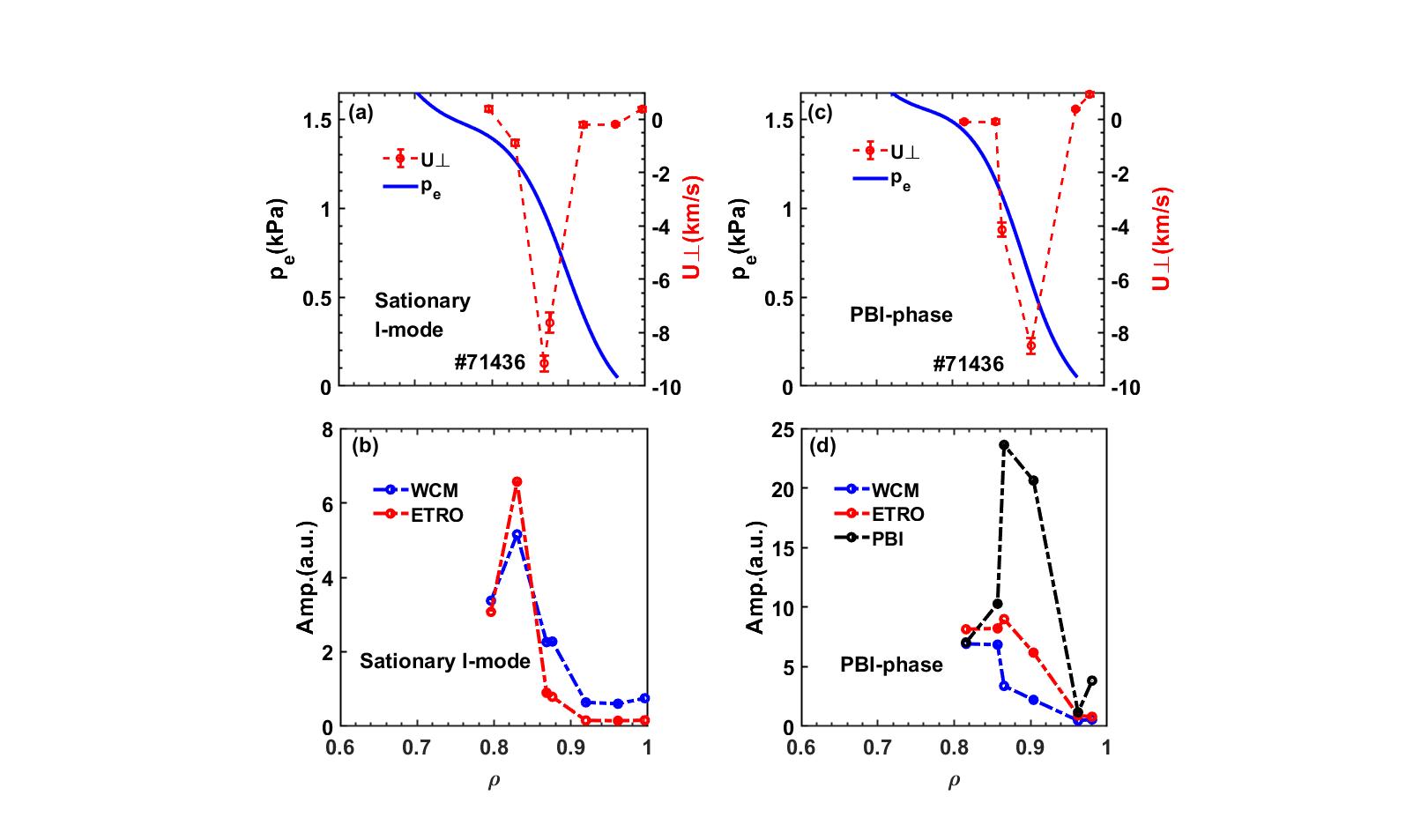}
\caption{(a), (c) {Pressure} profile and perpendicular velocity $U_{\perp}$ profile during the stationary I-mode and the PBI-phase in shot 71436 respectively. (b) Profile of the WCM amplitude and the ETRO amplitude during the stationary I-mode. (d) Profile of theWCM amplitude, the ETRO amplitude, and {the phase perturbation caused by PBI } during the PBI-phase.}
\label{PBIprofile}
\end{figure}

The radial distribution of the WCM and the ETRO during the stationary I-mode and the PBI-phase could be estimated through the phase derivative power density spectra from the multi-channel DR measurement \cite{Zhou2013RSI}, as shown in Fig. \ref{PBIprofile}b and Fig. \ref{PBIprofile}d, respectively.
The {pressure} profile and turbulence  perpendicular velocity distribution are also displayed in Fig. \ref{PBIprofile}a and Fig. \ref{PBIprofile}c.
In this shot the integrating frequency range is from $40 - 100 kHz$ for the WCM and $9 - 11 kHz$ for the ETRO.
Consistent with previous results \cite{Liu_2020}, both the WCM and the ETRO reach the maximum intensity at the inner side of the electric field well during the stationary I-mode, and the feature still exists during the PBI-phase (Fig. \ref{PBIprofile}d). 
{The radial distribution of the phase perturbation caused by PBI is shown in Fig. \ref{PBIprofile}d. 
It can be found that PBI is localized in the pedestal.}

\subsection{Evolution of density profile during PBI-phase}

\begin{figure}[htbp]
\centering
\includegraphics[width=4.5in]{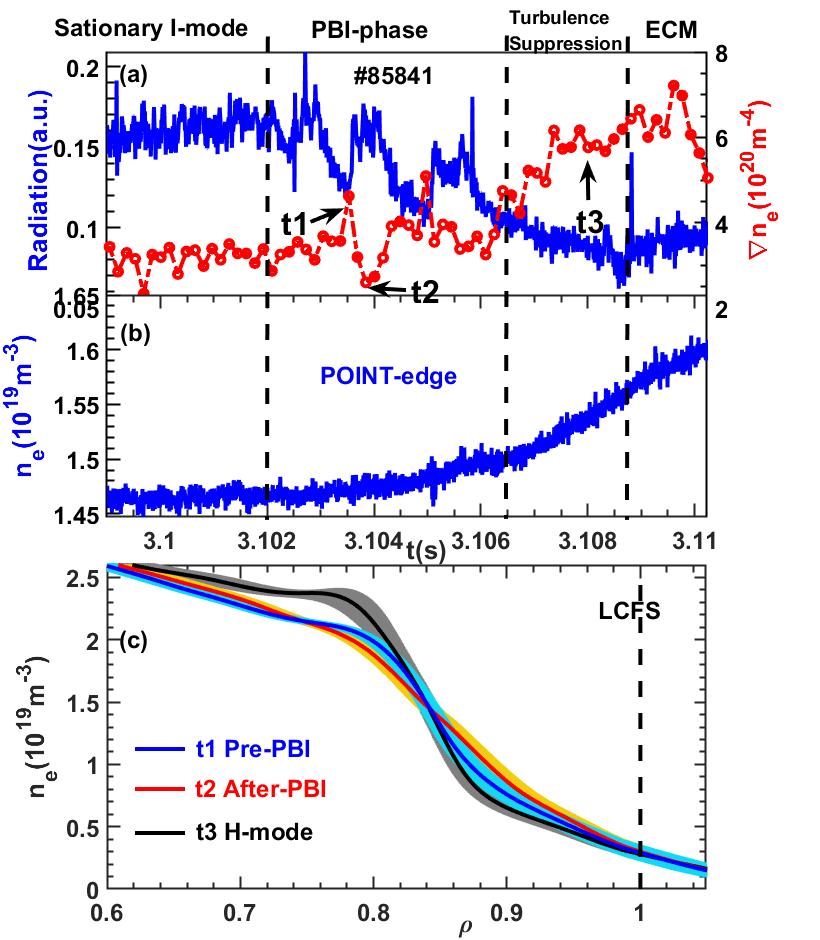}
\caption{(a) Blue solid line corresponds to the edge radiation signal from Bolometer in shot 85841, and red dotted line corresponds to the density gradient evolution in the pedestal region ($\rho = 0.84$). {(b) Chord-averaged density evolution at the edge during PBI-phase using POINT.} ({c}) Density profile at {pre-PBI ($t1$), after-PBI ($t2$), and H-mode ($t3$), respectively.} }
\label{ne85841}
\end{figure}
It is difficult to obtain the electron temperature evolutions during the PBI-phase because the ECE measurement is strongly interfered by the fast electrons generated by LHCD, which is the main auxiliary heating in EAST, especially in the plasma establishment phase.
Fig. \ref{ne85841} shows the density profile evolution during the PBI-phase in shot 85841, as well as the evolution of the density gradient in the pedestal region (at $\rho = 0.84$) , with the radiation signal from Bolometer as reference.
It should be noticed that this position ($\rho = 0.84$) has the maximum density gradient in H-mode.
{The density profile is measured with the time resolution of $\sim 0.16 ms$ by two Frequency-Modulated Continuous-Wave reflectometers, one is $50 – 75 GHz$ (V-band) with X-mode and the other is $33 – 50 GHz$ (Q-band) with O-mode \cite{JQHu2017}.
In addition, the zero-density position is directly estimated from the V-band X-mode reflectometer, and no initial guess is used.}
{And the shadow in Fig. \ref{ne85841}c shows the error bar for each density profile.}
The duration of the PBI-phase in this shot is about $5 ms$, from 3.102 $s$ to 3.107 $s$, with two distinct large bursts in the Bolometer signal.
The profiles before ($t = t1$) and after ($t = t2$) the first burst show that the PBI is actually a crash near the temperature pedestal top {($\rho \sim 0.81$)}, where the density gradient is significantly increased prior to the burst and then reduces after the burst, {while the profile during turbulence suppression phase (H-mode) is also displayed for comparison.}
{And the relative density perturbation at the pedestal top caused by PBI can be calculated using $\Delta n_e\_ped / n_e\_ped$ from reflectometer.
In this PBI, the value is about $6.2 \%$.
In addition, statistics on 32 PBIs show that the maxiumum value of the relative density perturbation caused by PBI could reach $8\%$, which is much smaller than that of large ELMs.
Similar results can be found in AUG, where the value is about $7\%$ \cite{Silvagni_2020}.}
{As shown in Fig. 5, both chord-averaged density at the edge and density gradient in the pedestal are significantly increased during the PBI-phase, implying that the PBI-phase is a gradual process of density pedestal establishment.}

\begin{figure}[htbp]
\centering
\includegraphics[width=6.4in]{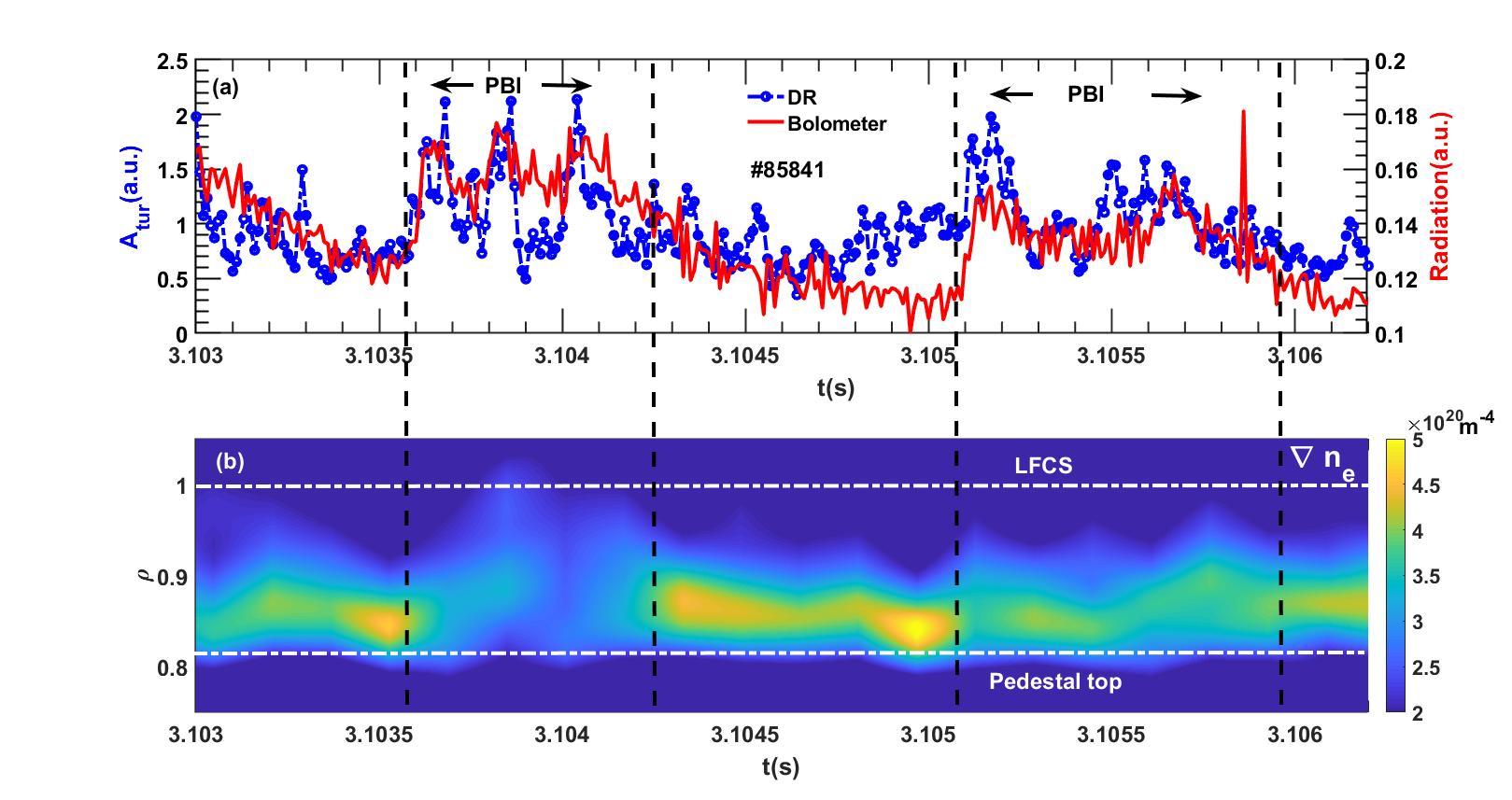}
\caption{(a) Blue line corresponds to turbulence intensity {(with the frequency range of $3 kHz - 2 MHz$)} evolution from DR  at $\rho \sim 0.84$,  red line corresponds to radiation evolution from Bolometer edge channel. (b) $\nabla n_e$ evolution in the pedestal region }
\label{gradne85841}
\end{figure}

The turbulence intensity {with the intergration frequency range of  $3 kHz - 2 MHz$} near the pedestal {is} shown in Fig. \ref{gradne85841}a,  suggesting that the turbulence could play an important role in triggering the PBIs.
To confirm the connection between the density gradient and turbulence evolutions, the temporal evolution of turbulence intensity, as well as the $\nabla n_e$ near the pedestal  are displayed in Fig. \ref{gradne85841}, respectively.
The intervals with significant increase of turbulence intensity are marked by the black dashed line in Fig. \ref{gradne85841}a, where the radiation power signal is also plotted as reference.
It can be found that prior to each PBI, a significant increase of the local density gradient can be clearly distinguished in Fig. \ref{gradne85841}b.

\subsection{Pedestal normalized density gradient triggering PBIs}

\begin{figure}[htbp]
\centering
\includegraphics[width=6.4in]{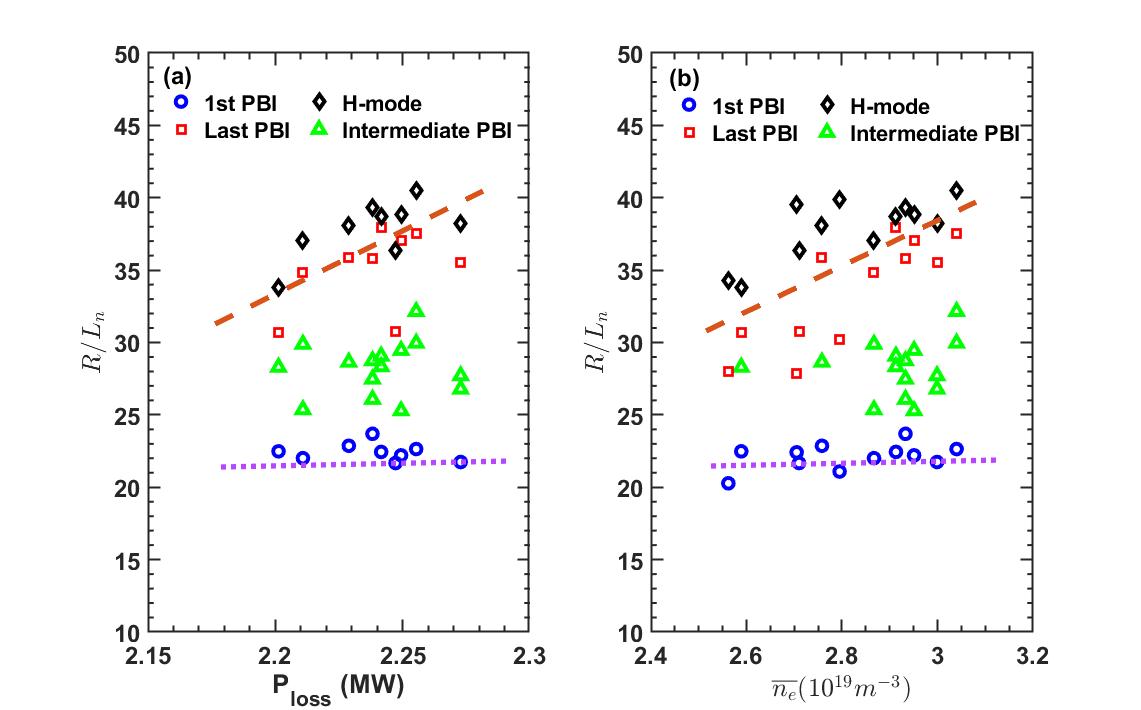}
\caption{Pedestal normalized density gradient, versus the loss power (a) and the chord-averaged density (b), respectively.}
\label{RLnplossne}
\end{figure}

\begin{table}
 \begin{center}
 \begin{tabular}{ |c|c|}
\hline
 Discharges & 9 \\
\hline
 Configuration & Unfavorable, LSN \\
\hline
$Ip (kA)$&500 \\
\hline
$n_e(10^{19}m^{-3})$&$2.5 - 3.1$ \\
\hline
$P_{heating}(MW)$ &$2.8 - 3.2$ \\
\hline
$W_{MHD}(kJ)$ & $100 - 140$ \\
\hline
$q_{95}$&$5.7 - 5.9$\\
\hline
 \end{tabular}
 \end{center}
 \caption{Experimental main parameters in Fig. \ref{RLnplossne}}
\label{IHRLn}
 \end{table}

Based on the I-H transition discharges with high temporal resolution density profile measurement, the statistic analysis is done to find out the possible threshold of density gradient triggering the PBIs.
Fig. \ref{RLnplossne} shows the pedestal normalized density gradient $R/L_n$ {, which is obtained using $R/L_n = R * (\nabla n_e\_ped/ n_e\_ped)_{max}$,} triggering the PBIs and the H-mode as a function of the loss power and the chord-averaged density, respectively.
{And the experimental main parameters of Fig. \ref{RLnplossne} are shown in the table \ref{IHRLn}.}
{The loss power is obtained using $P_{loss} = P_{heating} - dW_{dia}/dt - P_{rad}$, where $P_{heating}$ is the heating power, $dW_{dia}/dt$ is the time derivative of the stored energy, and $P_{rad}$ is the  radiation power.}
{It can be found that $R/L_n$ triggering the first PBI has a threshold, mostly in the range of $22 - 24$.
This could be consistent with a PBI triggering instability driven by the density gradient.}
And $R/L_n$ triggering the last PBI is about $30 - 40$.
The pedestal normalized density gradient at the moment, when the plasma enters H-mode, is slightly higher than that triggering the last PBI.
One interesting point is that $R/L_n$ triggering the last PBI seems to be proportional to the loss power and the chord-averaged density.
Due to the narrow range of the loss power, this conclusion is not definite.
The area can be divided into 3 parts by the brown dashed line and the purple dotted line {(both obtained from an experimental linear fitting)} in Fig. \ref{RLnplossne}.
H-mode is triggered when $R/L_n$ is in the region above the brown dashed line.
{If the value of the pedestal normalized density gradient is too small and does not reach the threshold, then the PBI cannot be excited.}

\subsection{Discussions}

\begin{figure}[htbp]
\centering
\includegraphics[width=6in]{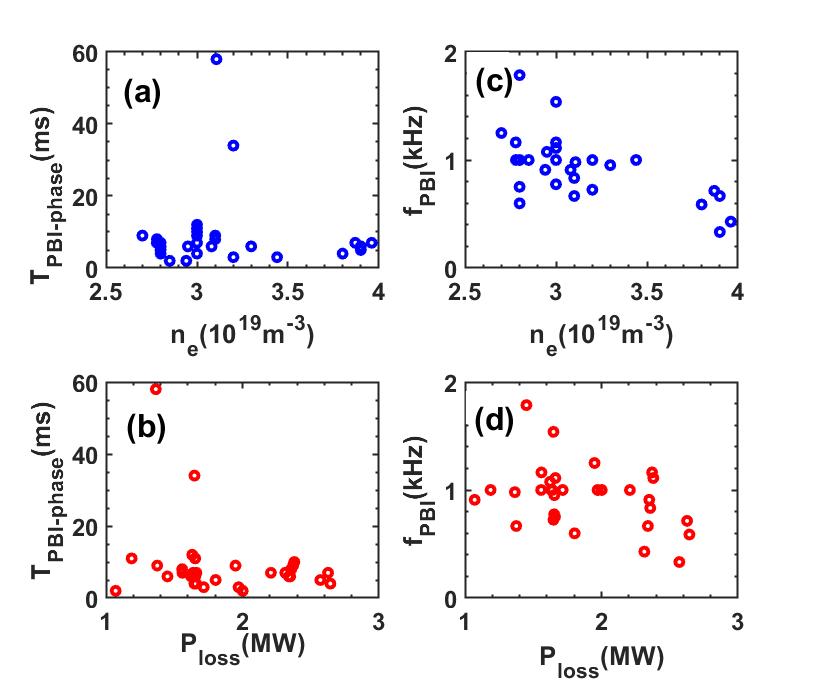}
\caption{(a) Chord-averaged density versus the duration of PBI-phase. (b) Loss power versus the duration of PBI-phase. (c) Chord-averaged density versus {the PBI frequency}. (d) Loss power versus {the PBI frequency}.}
\label{statistics}
\end{figure}

Based on the current discharges with I-H transition, the statistic results of the PBI-phase duration and PBI frequency are shown in Fig. \ref{statistics}a and Fig. \ref{statistics}b.
The PBI-phase duration is varied from $2$ to $58 ms$, with the average value of $7.5 ms$, and no obvious relationship could be found between the duration and the chord-averaged density or the loss power.
And the frequency of PBI $f_{PBI}$, which is calculated by the inverse of the time difference of two adjacent PBI onset, with the bolometer signal as reference, is varied from $0.5$ to $2 kHz$.
The most interesting point is that the PBI frequency seems to be inversely proportional to the density and the loss power.
The possible relationship between the pedestal plasma parameters and the PBI frequency will be further investigated after enough database of I-H transition without LHCD.

\begin{figure}[htbp]
\centering
\subfigure{
\label{Fig. sub.1}
\includegraphics[width=6in]{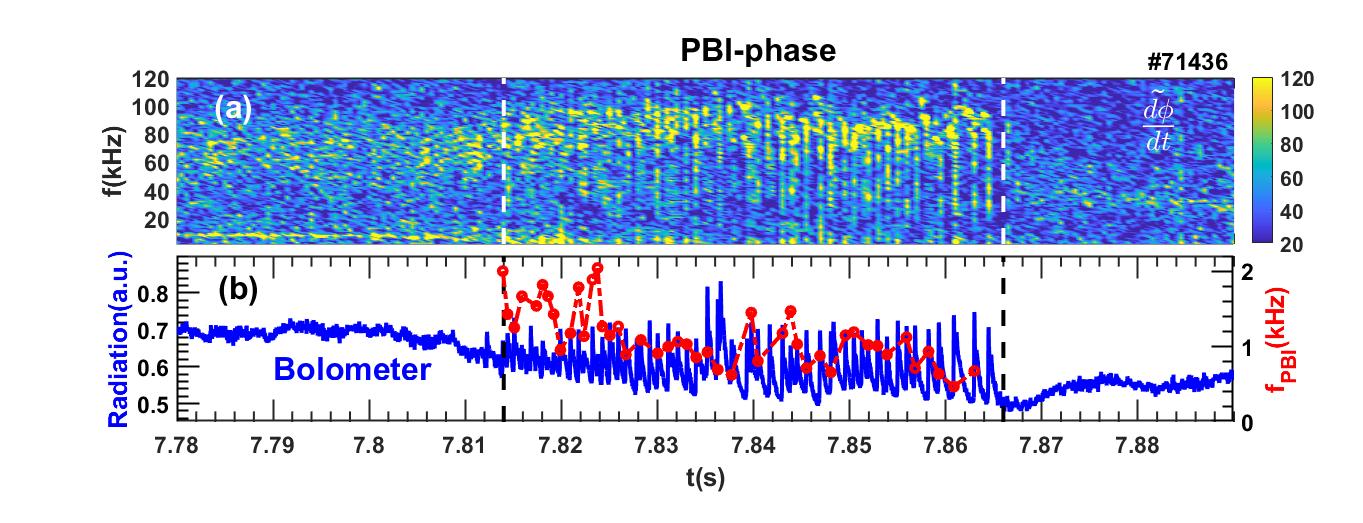}
}
\subfigure{
\label{Fig. sub.2}
\includegraphics[width=6in]{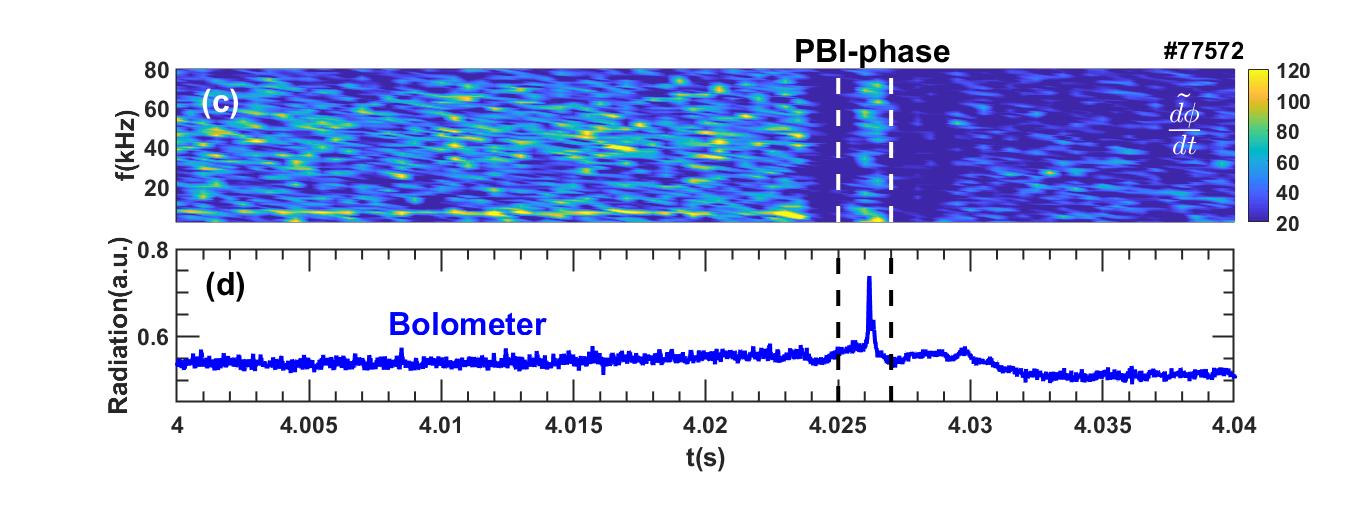}
}
\subfigure{
\label{Fig. sub.3}
\includegraphics[width=6in]{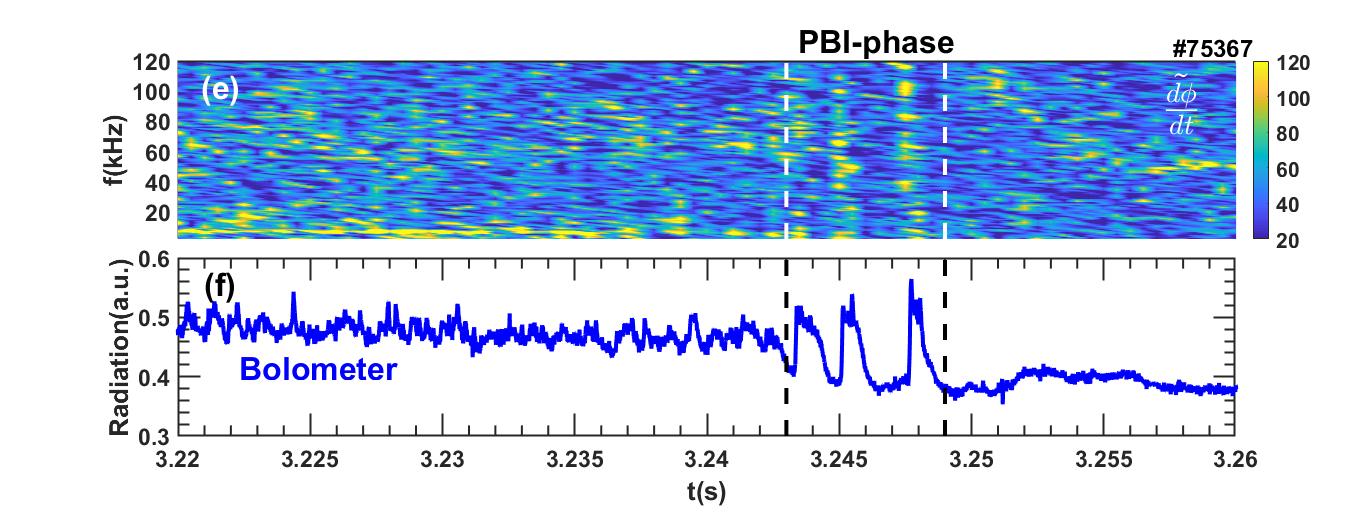}
}
\caption{(a), (c), (e) Time-frequency spectra of DR $d\phi/dt$ at the shot 71436, 77572, 75367.  (b), (d), (f) Radiation evolution of Bolometer at the shot 71436, 77572, 75367. {(b) Red line corresponds to the evolution of the PBI frequency.}}
\label{PBItype}
\end{figure}

{It should be noticed that PBI-phases are varied, as shown in Fig. \ref{PBItype}.}
The duration of the PBI-phase could last from several milliseconds to dozens of milliseconds.
{The long duration of PBI-phase is likely due to the slow growth of the pedestal normalized density gradient}
{It can be found that the PBI frequency seems to decrease gradually from $2 kHz$ to $0.5 kHz$, as shown in Fig. \ref{PBItype}a }
In addition, the PBI-phase with only one burst is displayed in Fig. \ref{PBItype}\st{b}{c and d}.

 \begin{table}
 \begin{center}
 \begin{tabular}{ |c|c|c|}
\hline
   & PBIs &PREs  \\
\hline
 Appearance  & I-H transition & I-H transition \\
\hline
Precursor & $\nabla n_e$ &$\delta B$ \\
\hline
Drive   &$\nabla n_e$  & Sawtooth (C-Mod); $\delta B$ \\
\hline
Frequency ($kHz$)   &$0.5 - 2$  & $0.1 - 0.7$ \\
\hline
Energy loss   & $1 \%$ &$2 - 3 \%$  \\
\hline
 \end{tabular}
 \end{center}
 \caption{Comparison between PBIs and PREs}
\label{comparison_PBIs_PREs}
 \end{table}
 
{The comparison between PBIs and PREs is shown in table \ref{comparison_PBIs_PREs}.
In AUG and C-Mod, PREs with the frequency range of $0.1 - 0.7 kHz$ mainly occur during the I-H transition, which typically result in  relative energy loss of $2 -3\%$ \cite{Silvagni2021NF, Bonanomi2021NF}. 
Moreover, PREs can be triggered by the core sawtooth instabilities in C-Mod.  
In addition, a magnetic fluctuation close to the separatrix is observed before the PRE onset in AUG and C-Mod. 
These events were simulated in \cite{Manz2021POP}.
The simulation shows that magnetic fluctuations are induced at higher plasma beta and excite the interchange mode.
Then the interchange mode leads to a steeper gradient across the pedestal region, which provokes the PREs.
In EAST, PBIs appear during the I-H transition, with the frequency range of $0.5 - 2 kHz$. 
And the relative loss of enengy is around $1\%$. 
Prior to each PBI, a significant increase of density gradient close to the pedestal can be distinguished. 
Moreover, there is a threshold of the pedestal normalized density gradient for the first PBI onset.
This could be consistent with a PBI triggering instability driven by the density gradient.
In short, although PBIs and PREs are similar in appearance, frequency, and the relative loss of energy, the precursor and drive of PBIs are significantly different from those of PREs.
 }

\section{Conclusions}

Pedestal burst instabilities (PBIs), featuring alternative turbulence suppression and bursts in the pedestal, can be clearly observed by most of the edge diagnostics (such as DR, Bolometer, SXR, ECE, Mirnov Probe, $D_{\alpha}$, and Div-LPs) during the I-H transition in the EAST tokamak.
{The radial distribution of the phase perturbation caused by PBI shows that PBI is localized in the pedestal.}
Prior to each PBI, a significant increase of density gradient in the pedestal can be distinguished.
Then the turbulence burst is generated, accompanied by the relaxation of the density profile.
{The relative density perturbation caused by PBI is about $6 - 8\%$, which is much smaller than that of large ELMs.}
Considering that PBI is just intermittent turbulence driven process, both the deposition range and the intensity of the induced particle flux are much smaller than that caused by large ELMs.
{And the significant increase of the chord-averaged density and the density gradient during the PBI-phase implies that the PBI-phase is a gradual process of density pedestal establishment.}

Statistic analyses show that the pedestal normalized density gradient $R/L_n$ triggering the first PBI has a threshold value, mostly in the range of $22 - 24$.
{This threshold value suggests that a PBI triggering instability could be driven by the density gradient.}
And $R/L_n$ triggering the last PBI is about $30 - 40$.
The pedestal normalized density gradient at the moment, when the plasma enters H-mode, is slightly higher than that triggering the last PBI.
In addition, the frequency of PBIs is likely to be inversely proportional to the chord-averaged density and the loss power.

In summary, {PBI-phase is a gradual process of density pedestal establishment,} accompanied by quasi-periodic turbulence generation close to the pedestal region, as well as the outward particle fluxes and the transient relaxation of the edge profiles.
The appearance of PBIs and the pedestal normalized density gradient prompt increase prior to PBIs allow identifying the precursor for controlling I-H transition.
The next step of our research will focus on the role of temperature gradient on driving the turbulence bursts, as well as the turbulence generation mechanism and the possible relationship between the turbulence generation and the unfavorable/favorable configuration.

\section*{Acknowledgments:}
This work was supported by the Natural Science Foundation of China under Grant Nos. U1967206, 11975231 and 11922513, the National MCF Energy R$\&$D Program under Grant Nos. 2017YFE0301204 and 2018YFE0311200, the Users with Excellence Program of Hefei Science Center CAS under Grant No. 2020HSC-UE009 and Fundamental Research Funds for the Central Universities.  
We also acknowledge the EAST team for supporting the experiments.

\section*{Reference:}

\bibliographystyle{unsrt}
\bibliography{ref}

\begin{thebibliography}{10}

\bibitem{wagner1982}
Fritz Wagner, G~Becker, K~Behringer, D~Campbell, A~Eberhagen, W~Engelhardt,
  G~Fussmann, O~Gehre, J~Gernhardt, G~v Gierke, et~al.
\newblock Regime of improved confinement and high beta in neutral-beam-heated
  divertor discharges of the asdex tokamak.
\newblock {\em Physical Review Letters}, 49(19):1408, 1982.

\bibitem{wagner1984}
Fritz Wagner, G~Fussmann, T~Grave, M~Keilhacker, M~Kornherr, K~Lackner,
  K~McCormick, ER~M{\"u}ller, A~St{\"a}bler, G~Becker, et~al.
\newblock Development of an edge transport barrier at the h-mode transition of
  asdex.
\newblock {\em Physical Review Letters}, 53(15):1453, 1984.

\bibitem{rebut1995}
P-H Rebut et~al.
\newblock Iter: the first experimental fusion reactor.
\newblock {\em Fusion engineering and design}, 30(1-2):85--118, 1995.

\bibitem{zohm1996}
Hartmut Zohm.
\newblock Edge localized modes (elms).
\newblock {\em Plasma Physics and Controlled Fusion}, 38(2):105, 1996.

\bibitem{leonard2014edge}
Anthony~W Leonard.
\newblock Edge-localized-modes in tokamaks.
\newblock {\em Physics of Plasmas}, 21(9):090501, 2014.

\bibitem{evans2006edge}
Todd~E Evans, Richard~A Moyer, Keith~H Burrell, Max~E Fenstermacher, Ilon
  Joseph, Anthony~W Leonard, Thomas~H Osborne, Gary~D Porter, Michael~J
  Schaffer, Philip~B Snyder, et~al.
\newblock Edge stability and transport control with resonant magnetic
  perturbations in collisionless tokamak plasmas.
\newblock {\em nature physics}, 2(6):419--423, 2006.

\bibitem{milora1995pellet}
SL~Milora, WA~Houlberg, LL~Lengyel, and V~Mertens.
\newblock Pellet fuelling.
\newblock {\em Nuclear Fusion}, 35(6):657, 1995.

\bibitem{xiao2012elm}
WW~Xiao, PH~Diamond, XL~Zou, JQ~Dong, XT~Ding, LH~Yao, BB~Feng, CY~Chen,
  WL~Zhong, M~Xu, et~al.
\newblock Elm mitigation by supersonic molecular beam injection into the h-mode
  pedestal in the hl-2a tokamak.
\newblock {\em Nuclear Fusion}, 52(11):114027, 2012.

\bibitem{zheng2013comparison}
Xingwei Zheng, Jiangang Li, Jiansheng Hu, Jiahong Li, Rui Ding, Bin Cao, and
  Jinhua Wu.
\newblock Comparison between gas puffing and supersonic molecular beam
  injection in plasma density feedback experiments in east.
\newblock {\em Plasma Physics and Controlled Fusion}, 55(11):115010, 2013.

\bibitem{xiao2017effect}
GL~Xiao, WL~Zhong, XL~Zou, XR~Duan, AD~Liu, XY~Bai, J~Cheng, ZY~Cui, L~Delpech,
  XT~Ding, et~al.
\newblock Effect of lower hybrid current drive on pedestal instabilities in the
  hl-2a tokamak.
\newblock {\em Physics of Plasmas}, 24(12):122507, 2017.

\bibitem{zhang2018control}
YP~Zhang, D~Mazon, XL~Zou, WL~Zhong, JM~Gao, K~Zhang, P~Sun, CF~Dong, ZY~Cui,
  Yi~Liu, et~al.
\newblock Control of edge localized modes by pedestal deposited impurity in the
  hl-2a tokamak.
\newblock {\em Nuclear Fusion}, 58(4):046018, 2018.

\bibitem{burrell2001quiescent}
KH~Burrell, ME~Austin, DP~Brennan, JC~DeBoo, EJ~Doyle, C~Fenzi, C~Fuchs,
  P~Gohil, CM~Greenfield, RJ~Groebner, et~al.
\newblock Quiescent double barrier high-confinement mode plasmas in the diii-d
  tokamak.
\newblock {\em Physics of Plasmas}, 8(5):2153--2162, 2001.

\bibitem{greenwald1999characterization}
M~Greenwald, R~Boivin, P~Bonoli, R~Budny, C~Fiore, J~Goetz, R~Granetz,
  A~Hubbard, I~Hutchinson, J~Irby, et~al.
\newblock Characterization of enhanced d$\alpha$ high-confinement modes in
  alcator c-mod.
\newblock {\em Physics of Plasmas}, 6(5):1943--1949, 1999.

\bibitem{kamiya2003observation}
Kensaku Kamiya, H~Kimura, H~Ogawa, H~Kawashima, K~Tsuzuki, M~Sato, Y~Miura,
  JFT-2M group, et~al.
\newblock Observation of high recycling steady h-mode edge and compatibility
  with improved core confinement mode on jft-2m.
\newblock {\em Nuclear fusion}, 43(10):1214, 2003.

\bibitem{kamada2000disappearance}
Y~Kamada, T~Oikawa, L~Lao, T~Takizuka, T~Hatae, A~Isayama, J~Manickam,
  M~Okabayashi, T~Fukuda, and K~Tsuchiya.
\newblock Disappearance of giant elms and appearance of minute grassy elms in
  jt-60u high-triangularity discharges.
\newblock {\em Plasma Physics and Controlled Fusion}, 42(5A):A247, 2000.

\bibitem{stober2001type}
J~Stober, M~Maraschek, GD~Conway, O~Gruber, A~Herrmann, ACC Sips, W~Treutterer,
  H~Zohm, and ASDEX~Upgrade Team.
\newblock Type ii elmy h modes on asdex upgrade with good confinement at high
  density.
\newblock {\em Nuclear Fusion}, 41(9):1123, 2001.

\bibitem{maingi2005observation}
R~Maingi, K~Tritz, ED~Fredrickson, JE~Menard, SA~Sabbagh, D~Stutman, MG~Bell,
  RE~Bell, CE~Bush, DA~Gates, et~al.
\newblock Observation of a high performance operating regime with small
  edge-localized modes in the national spherical torus experiment.
\newblock {\em Nuclear fusion}, 45(4):264, 2005.

\bibitem{walk2014edge}
JR~Walk, JW~Hughes, AE~Hubbard, JL~Terry, DG~Whyte, AE~White, SG~Baek,
  ML~Reinke, C~Theiler, RM~Churchill, et~al.
\newblock Edge-localized mode avoidance and pedestal structure in i-mode
  plasmas.
\newblock {\em Physics of Plasmas}, 21(5):056103, 2014.

\bibitem{Happel2016PPCF}
T~Happel, P~Manz, F~Ryter, M~Bernert, M~Dunne, Pascale Hennequin,
  A~Hetzenecker, U~Stroth, GD~Conway, L~Guimarais, et~al.
\newblock The i-mode confinement regime at asdex upgrade: global properties and
  characterization of strongly intermittent density fluctuations.
\newblock {\em Plasma Physics and Controlled Fusion}, 59(1):014004, 2016.

\bibitem{whyte2010mode}
DG~Whyte, AE~Hubbard, JW~Hughes, B~Lipschultz, JE~Rice, ES~Marmar, M~Greenwald,
  I~Cziegler, A~Dominguez, T~Golfinopoulos, et~al.
\newblock I-mode: an h-mode energy confinement regime with l-mode particle
  transport in alcator c-mod.
\newblock {\em Nuclear Fusion}, 50(10):105005, 2010.

\bibitem{ryter2016mode}
F~Ryter, R~Fischer, JC~Fuchs, T~Happel, RM~McDermott, Eleonora Viezzer,
  E~Wolfrum, L~Barrera Orte, M~Bernert, A~Burckhart, et~al.
\newblock I-mode studies at asdex upgrade: Li and ih transitions, pedestal and
  confinement properties.
\newblock {\em Nuclear Fusion}, 57(1):016004, 2016.

\bibitem{liu2020power}
YJ~Liu, ZX~Liu, AD~Liu, C~Zhou, X~Feng, Y~Yang, T~Zhang, TY~Xia, HQ~Liu, MQ~Wu,
  et~al.
\newblock Power threshold and confinement of the i-mode in the east tokamak.
\newblock {\em Nuclear Fusion}, 60(8):082003, 2020.

\bibitem{greenwald1997h}
Martin Greenwald, RL~Boivin, F~Bombarda, PT~Bonoli, CL~Fiore, D~Garnier,
  JA~Goetz, SN~Golovato, MA~Graf, RS~Granetz, et~al.
\newblock H mode confinement in alcator c-mod.
\newblock {\em Nuclear Fusion}, 37(6):793, 1997.

\bibitem{Ryter_1998}
F~Ryter, W~Suttrop, B~Brüsehaber, M~Kaufmann, V~Mertens, H~Murmann, A~G
  Peeters, J~Stober, J~Schweinzer, H~Zohm, and ASDEX~Upgrade Team.
\newblock H-mode power threshold and transition in {ASDEX} upgrade.
\newblock {\em Plasma Physics and Controlled Fusion}, 40(5):725--729, may 1998.

\bibitem{manz2015geodesic}
Peter Manz, Ph~Lauber, VE~Nikolaeva, T~Happel, F~Ryter, G~Birkenmeier, Anton
  Bogomolov, GD~Conway, ME~Manso, M~Maraschek, et~al.
\newblock Geodesic oscillations and the weakly coherent mode in the i-mode of
  asdex upgrade.
\newblock {\em Nuclear Fusion}, 55(8):083004, 2015.

\bibitem{Manz2020NF}
P.~Manz, T.~Happel, U.~Stroth, T.~Eich, and D.~Silvagni and.
\newblock Physical mechanism behind and access to the i-mode confinement regime
  in tokamaks.
\newblock {\em Nuclear Fusion}, 60(9):096011, aug 2020.

\bibitem{zxliu2016pop}
Z.~X. Liu, X.~Q. Xu, X.~Gao, A.~E. Hubbard, J.~W. Hughes, J.~R. Walk,
  C.~Theiler, T.~Y. Xia, S.~G. Baek, T.~Golfinopoulos, D.~Whyte, T.~Zhang, and
  J.~G. Li.
\newblock The physics mechanisms of the weakly coherent mode in the alcator
  c-mod tokamak.
\newblock {\em Physics of Plasmas}, 23(12):120703, 2016.

\bibitem{Cziegler2017PRL}
I.~Cziegler, A.~E. Hubbard, J.~W. Hughes, J.~L. Terry, and G.~R. Tynan.
\newblock Turbulence nonlinearities shed light on geometric asymmetry in
  tokamak confinement transitions.
\newblock {\em Phys. Rev. Lett.}, 118:105003, Mar 2017.

\bibitem{Liu_2020}
A.D. Liu, X.L. Zou, M.K. Han, T.B. Wang, C.~Zhou, M.Y. Wang, Y.M. Duan,
  G.~Verdoolaege, J.Q. Dong, Z.X. Wang, F.~Xi, J.L. Xie, G.~Zhuang, W.X. Ding,
  S.B. Zhang, Y.~Liu, H.Q. Liu, L.~Wang, Y.Y. Li, Y.M. Wang, B.~Lv, G.H. Hu,
  Q.~Zhang, S.X. Wang, H.L. Zhao, C.M. Qu, Z.X. Liu, Z.Y. Liu, J.~Zhang, J.X.
  Ji, X.M. Zhong, T.~Lan, H.~Li, W.Z. Mao, and W.D. Liu.
\newblock Experimental identification of edge temperature ring oscillation and
  alternating turbulence transitions near the pedestal top for sustaining
  stationary i-mode.
\newblock {\em Nuclear Fusion}, 60(12):126016, sep 2020.

\bibitem{feng2019mode}
X~Feng, AD~Liu, C~Zhou, ZX~Liu, MY~Wang, G~Zhuang, XL~Zou, TB~Wang, YZ~Zhang,
  JL~Xie, et~al.
\newblock I-mode investigation on the experimental advanced superconducting
  tokamak.
\newblock {\em Nuclear Fusion}, 59(9):096025, 2019.

\bibitem{Silvagni2021NF}
Davide Silvagni, Jim~L Terry, William McCarthy, Amanda~E Hubbard, Thomas Eich,
  Michael Faitsch, Luís Gil, Theodore Golfinopoulos, Gustavo Grenfell, Michael
  Griener, Tim Happel, Jerry~W Hughes, Ulrich Stroth, and Eleonora Viezzer.
\newblock I-mode pedestal relaxation events in the alcator c-mod and asdex
  upgrade tokamaks.
\newblock {\em Nuclear Fusion}, 2021.

\bibitem{hubbard2016multi}
AE~Hubbard, T~Osborne, F~Ryter, M~Austin, L~Barrera Orte, RM~Churchill,
  I~Cziegler, M~Fenstermacher, R~Fischer, S~Gerhardt, et~al.
\newblock Multi-device studies of pedestal physics and confinement in the
  i-mode regime.
\newblock {\em Nuclear Fusion}, 56(8):086003, 2016.

\bibitem{Silvagni_2020}
D.~Silvagni, T.~Eich, T.~Happel, G.F. Harrer, M.~Griener, M.~Dunne, M.~Cavedon,
  M.~Faitsch, L.~Gil, D.~Nille, B.~Tal, R.~Fischer, U.~Stroth, D.~Brida,
  P.~David, P.~Manz, E.~Viezzer, and and.
\newblock I-mode pedestal relaxation events at {ASDEX} upgrade.
\newblock {\em Nuclear Fusion}, 60(12):126028, oct 2020.

\bibitem{wu2007overview}
Songtao Wu, EAST team, et~al.
\newblock An overview of the east project.
\newblock {\em Fusion Engineering and Design}, 82(5-14):463--471, 2007.

\bibitem{wan2017overview}
BN~Wan, YF~Liang, XZ~Gong, JG~Li, N~Xiang, GS~Xu, YW~Sun, L~Wang, JP~Qian,
  HQ~Liu, et~al.
\newblock Overview of east experiments on the development of high-performance
  steady-state scenario.
\newblock {\em Nuclear Fusion}, 57(10):102019, 2017.

\bibitem{li2013long}
J~Li, HY~Guo, BN~Wan, XZ~Gong, YF~Liang, GS~Xu, KF~Gan, JS~Hu, HQ~Wang, L~Wang,
  et~al.
\newblock A long-pulse high-confinement plasma regime in the experimental
  advanced superconducting tokamak.
\newblock {\em Nature physics}, 9(12):817--821, 2013.

\bibitem{Zhou2013RSI}
C.~Zhou, A.~D. Liu, X.~H. Zhang, J.~Q. Hu, M.~Y. Wang, H.~Li, T.~Lan, J.~L.
  Xie, X.~Sun, W.~X. Ding, W.~D. Liu, and C.~X. Yu.
\newblock Microwave doppler reflectometer system in the experimental advanced
  superconducting tokamak.
\newblock {\em Review of Scientific Instruments}, 84(10):103511, 2013.

\bibitem{zou1999eps}
XL~Zou, TF~Seak, M~Paume, JM~Chareau, C~Bottereau, and G~Leclert.
\newblock Poloidal rotation measurement in tore supra by oblique reflectometry.
\newblock In {\em Proc. 26th EPS Conf. on Controlled Fusion and Plasma
  Physics}, volume 23J, page 1041, 1999.

\bibitem{hillesheim2010rsi}
JC~Hillesheim, WA~Peebles, TL~Rhodes, L~Schmitz, AE~White, and TA~Carter.
\newblock New plasma measurements with a multichannel millimeter-wave
  fluctuation diagnostic system in the diii-d tokamak.
\newblock {\em Review of Scientific Instruments}, 81(10):10D907, 2010.

\bibitem{fanack1996ppcf}
C~Fanack, I~Boucher, F~Clairet, S~Heuraux, G~Leclert, and XL~Zou.
\newblock Ordinary-mode reflectometry: modification of the scattering and
  cut-off responses due to the shape of localized density fluctuations.
\newblock {\em Plasma physics and controlled fusion}, 38(11):1915, 1996.

\bibitem{Zhong2016ppcf}
W~L Zhong, X~L Zou, Z~B Shi, X~R Duan, Y~Xu, M~Xu, W~Chen, M~Jiang, Z~C Yang,
  B~Y Zhang, P~W Shi, Z~T Liu, X~M Song, J~Cheng, X~Q Ji, Y~Zhou, D~L Yu, J~X
  Li, J~Q Dong, X~T Ding, Y~Liu, L~W Yan, Q~W Yang, and Y~Liu and.
\newblock Excitation of edge plasma instabilities and their role in pedestal
  saturation in the {HL}-2a tokamak.
\newblock 58(6):065001, apr 2016.

\bibitem{SHEN20132830}
Jie Shen, Yinxian Jie, Haiqing Liu, Xuechao Wei, Zhengxing Wang, and Xiang Gao.
\newblock Improved density measurement by fir laser interferometer on east
  tokamak.
\newblock {\em Fusion Engineering and Design}, 88(11):2830--2834, 2013.

\bibitem{Liu2014RSI}
H.~Q. Liu, Y.~X. Jie, W.~X. Ding, D.~L. Brower, Z.~Y. Zou, W.~M. Li, Z.~X.
  Wang, J.~P. Qian, Y.~Yang, L.~Zeng, T.~Lan, X.~C. Wei, G.~S. Li, L.~Q. Hu,
  and B.~N. Wan.
\newblock Faraday-effect polarimeter-interferometer system for current density
  measurement on east.
\newblock {\em Review of Scientific Instruments}, 85(11):11D405, 2014.

\bibitem{Xiang2018RSI}
H.~M. Xiang, T.~Zhang, F.~Wen, H.~Qu, M.~F. Wu, K.~N. Geng, G.~S. Li, Y.~M.
  Wang, X.~Han, Zi~X. Liu, F.~B. Zhong, K.~X. Ye, S.~B. Zhang, and X.~Gao.
\newblock Development of an ordinary mode multi-channel correlation
  reflectometer on east tokamak.
\newblock {\em Review of Scientific Instruments}, 89(10):10H103, 2018.

\bibitem{LIU201872}
Yong Liu, Hailin Zhao, Tianfu Zhou, Xiang Liu, Zeying Zhu, Xiang Han, Stefan
  Schmuck, John Fessey, Paul Trimble, C.W. Domier, N.C. Luhmann, Ang Ti,
  Erzhong Li, Bili Ling, Liqun Hu, Xi~Feng, Ahdi Liu, W.L. Rowan, H.~Huang, and
  P.E. Phillips.
\newblock Overview of the electron cyclotron emission measurements on east.
\newblock {\em Fusion Engineering and Design}, 136:72--75, 2018.
\newblock Special Issue: Proceedings of the 13th International Symposium on
  Fusion Nuclear Technology (ISFNT-13).

\bibitem{qing2010development}
Zang Qing, Zhao Junyu, Yang Li, Hu~Qingsheng, Jia Yanqing, Zhang Tao,
  Xi~Xiaoqi, SH~Bhatti, and Gao Xiang.
\newblock Development of a thomson scattering diagnostic system on east.
\newblock {\em Plasma Science and Technology}, 12(2):144, 2010.

\bibitem{duan2012resistive}
YM~Duan, LQ~Hu, ST~Mao, KY~Chen, SY~Lin, and EAST~Diagnostics Team.
\newblock The resistive bolometer for radiated power measurement on east.
\newblock {\em Review of Scientific Instruments}, 83(9):093501, 2012.

\bibitem{chen20162}
Kaiyun Chen, Liqing Xu, Liqun Hu, Yanmin Duan, Xueqin Li, Yi~Yuan, Songtao Mao,
  Xiuli Sheng, and Jinlong Zhao.
\newblock 2-d soft x-ray arrays in the east.
\newblock {\em Review of Scientific Instruments}, 87(6):063504, 2016.

\bibitem{Xu2016RSI}
J.~C. Xu, L.~Wang, G.~S. Xu, G.~N. Luo, D.~M. Yao, Q.~Li, L.~Cao, L.~Chen,
  W.~Zhang, S.~C. Liu, H.~Q. Wang, M.~N. Jia, W.~Feng, G.~Z. Deng, L.~Q. Hu,
  B.~N. Wan, J.~Li, Y.~W. Sun, and H.~Y. Guo.
\newblock Upgrade of langmuir probe diagnostic in iter-like tungsten mono-block
  divertor on experimental advanced superconducting tokamak.
\newblock {\em Review of Scientific Instruments}, 87(8):083504, 2016.

\bibitem{JI_2021}
Jiaxu JI, Adi LIU, Chu ZHOU, Xi~FENG, Shouxin WANG, Haiqing LIU, Hailin ZHAO,
  Yong LIU, Jin ZHANG, Zhaoyang LIU, Xiaoming ZHONG, Hongrui FAN, Ge~ZHUANG,
  Jinlin XIE, Tao LAN, Wenzhe MAO, Weixing DING, Hong LI, Zixi LIU, and Wandong
  LIU.
\newblock The investigation of quasi coherent mode on {EAST} using doppler
  reflectometry.
\newblock {\em Plasma Science and Technology}, 23(9):095106, jul 2021.

\bibitem{JQHu2017}
Hu~Jianqiang, LIU Ahdi, ZHOU Chu, Xiaohui ZHANG, WANG Mingyuan, Jin ZHANG, FENG
  Xi, LI~Hong, XIE Jinlin, LIU Wandong, et~al.
\newblock An accurate automated technique for quasi-optics measurement of the
  microwave diagnostics for fusion plasma.
\newblock {\em Plasma Science and Technology}, 19(8):084002, 2017.

\bibitem{Bonanomi2021NF}
N.~Bonanomi, C.~Angioni, D.~Silvagni, T.~Happel, U.~Plank, L.~Gil, P.A.
  Schneider, T.~Puetterich, the EUROFusion MST1~Team, and the ASDEX
  Upgrade~Team.
\newblock I-mode in non-deuterium plasmas in {ASDEX} upgrade.
\newblock {\em Nuclear Fusion}, 61(5):054001, apr 2021.

\bibitem{Manz2021POP}
P.~Manz, D.~Silvagni, O.~Grover, T.~Happel, T.~Eich, and M.~Griener.
\newblock Gyrofluid simulation of an i-mode pedestal relaxation event.
\newblock {\em Physics of Plasmas}, 28(10):102502, 2021.

\end{thebibliography}
\end{document}